%% file: ms.tex
\documentclass[10pt, conference, compsocconf]{IEEEtran}
\usepackage{mathptmx}

\usepackage{graphicx}
\usepackage{times}
\usepackage[portuguese]{babel}
\usepackage{tabularx}
\usepackage[utf8]{inputenc}
\let\labelindent\relax
\usepackage[inline]{enumitem}
\usepackage[autostyle]{csquotes}
\usepackage{color}
\usepackage{hyperref}

\usepackage{cite}
\begin{document}
%
\title{Levantamento de Requisitos para Jogos Educativos Infantis}


\author{\IEEEauthorblockN{Clara Andrade Pimentel}
\IEEEauthorblockA{Universidade Federal de Minas Gerais\\
Belo Horizonte, Minas Gerais\\
Email: clarapimentel@ufmg.br}
\and
\IEEEauthorblockN{Philipe de Freitas Melo}
\IEEEauthorblockA{Universidade Federal de Minas Gerais\\
Belo Horizonte, Minas Gerais\\
Email: philipe@dcc.ufmg.br}
\and
\IEEEauthorblockN{Marília Lyra Bergamo}
\IEEEauthorblockA{Universidade Federal de Minas Gerais\\
Belo Horizonte, Minas Gerais\\
Email: marilialb@eba.ufmg.br}
}
\maketitle

\begin{abstract}
O uso de jogos dentro de sala de aula tem crescido atualmente, contudo, ainda faltam soluções melhores para problemas como a conciliação do lazer com aula ou a falta de atenção das crianças. Neste trabalho são demonstradas as metodologias utilizadas para o levantamento e análise de requisitos para o desenvolvimento de jogos educativos infantis, com o público alvo de crianças de 5 a 9 anos de idade. É levantada uma lista de requisitos base que podem melhorar o desenvolvimento dos jogos educativos e promover uma base para o desenvolvimento posterior de um jogo. É possível dar continuidade ao trabalho abordando teorias da educação, o processo de gamificação da educação e outros métodos de levantamento de requisitos.

\end{abstract}


%
\IEEEpeerreviewmaketitle

\section{Introdução}
\input{introduction.tex}

\section{Trabalhos Relacionados}
\input{related_work.tex}

\section{Desenvolvimento} 
\input{research_process.tex}

\subsection{Análise de Jogos Existentes}
\input{others_ideas.tex}

\section{Resultados} \label{sec:resultados}
\input{results.tex}

\section{Conclusão e Trabalhos Futuros} \label{sec:conclusao}
\input{conclusion.tex}

\bibliographystyle{IEEEtran}
\bibliography{template.bib}

\end{document}

%% file: introduction.tex
Jogos têm sido uma forma de entretenimento, diversão e passatempo há anos na vida das pessoas. Desde os jogos esportivos, os jogos de tabuleiro como Monopoly, xadrez, até os jogos digitais, todos têm um papel importante em nosso cotidiano. Familiares e amigos se reúnem para assistir jogos e jogar juntos, as pessoas compram \textit{videogames} e computadores para se divertirem com os jogos do momento.

Desde o surgimento dos primeiros jogos digitais, a ideia de aplicá-los em outros segmentos, não só no lazer, ganhou força. A utilização de elementos lúdicos na educação, tais como a dança e música, já é difundida há muito tempo. Os jogos possuem uma capacidade única de criação de mundos e regras próprias, abrindo infinitas possibilidades de conteúdo e novas interações, que associado à atenção que eles captam dos jogadores (o nível de imersão que possuem), os tornam ferramentas lúdicas com um grande potencial de aplicação na educação \cite{de1998educaccao}.

Semelhante aos jogos esportivos, os jogos digitais e de tabuleiro possuem regras próprias e limites físicos ou imaginários que diferenciam o mundo do jogo do mundo real. Esses universos criados pelos jogos, seja o jogo de dama ou \textit{Minecraft} \footnote{\textit{Minecraft}. Mojang, 2009. Disponível em: \url{https://minecraft.net/en/}.}, são condicionados a um conjunto de leis que determinam o que acontece dentro daquele pequeno mundo, isso é o que chamamos de círculo mágico \cite{huizinga1971homo}. Os jogos de tabuleiro e digitais, diferentemente dos esportes, abrem maior espaço para a imaginação. Eles utilizam elementos narrativos para criar universos mais consistentes e imaginários, capazes de novos níveis de imersão dos jogadores.

Contudo, a utilização de jogos dentro da sala de aula é algo ainda distante da educação brasileira, devido à falta de acesso dos educadores às ferramentas, as dificuldades de desenvolvimento e os custos de implementação. Os jogos educativos também possuem dificuldades em atingir o público infanto-juvenil, pois existem complicações no equilíbrio do conteúdo didático com a diversão do jogo, o que ora afasta as crianças ora não é efetivo no ensino \cite{dondlinger2007}. Jogos ainda ficam à margem da educação, sendo reservado pouco tempo e frequência para a utilização deles dentro das salas de aula.

Apesar de existirem jogos criados para fins educativos, muitos deles caem em desuso rapidamente ou não são tão efetivos na educação o quanto poderiam ser. Como discutido por Druin (2002)~\cite{druin2002role} e Gros (2007)~\cite{gros2007}, existe uma categoria de jogos que surgiu primariamente no cenário dos jogos educativos que é marcada pela prática repetitiva de tarefas. Esses primeiros \textit{edutainment} games se provaram pouco eficazes na educação, não propiciando chance de um aprendizado contínuo. Jogos desse caráter ainda existem, onde a criança é colocada para repetir tarefas continuamente, como operações matemáticas, nomear itens, responder perguntas, com o objetivo de fixar o conteúdo didático.

Com a evolução das teorias educacionais e do processo de inserção da criança como agente participativo no desenvolvimento de novas tecnologias, os jogos evoluíram, ganhando novas caras. Hoje os jogos educativos têm foco no usuário e no contexto social, sendo utilizados jogos de simulação, aventura e outros gêneros que exploram não só o ensino, mas o crescimento pessoal e social do aluno. Entretanto, os fatores mais desafiadores na criação de jogos são ainda o balanceamento entre uma interface atraente e de fácil uso com a qualidade do conteúdo didático \cite{crozat1999method}.

\begin{figure}[h!]
  \centering
  \includegraphics[width=0.7\linewidth]{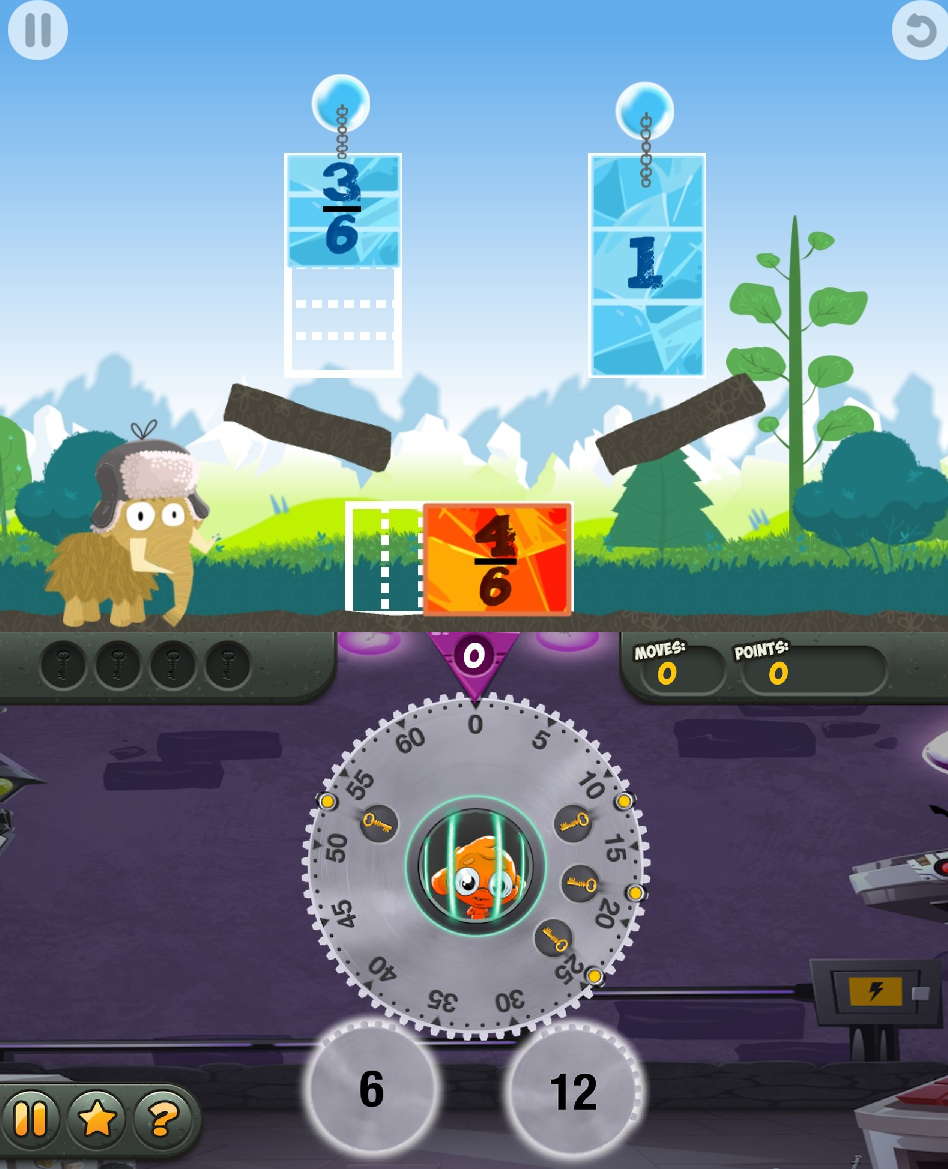}
  \caption{\textit{Screenshot de jogos com exercícios matemáticos. Acima Slice Fractions \cite{slicefractions} e abaixo Wuzzit Trouble \cite{wuzzit}}}
  \label{fig:slicefractions}
\end{figure}

\begin{figure}[h!]
  \centering
  \includegraphics[width=0.9\linewidth]{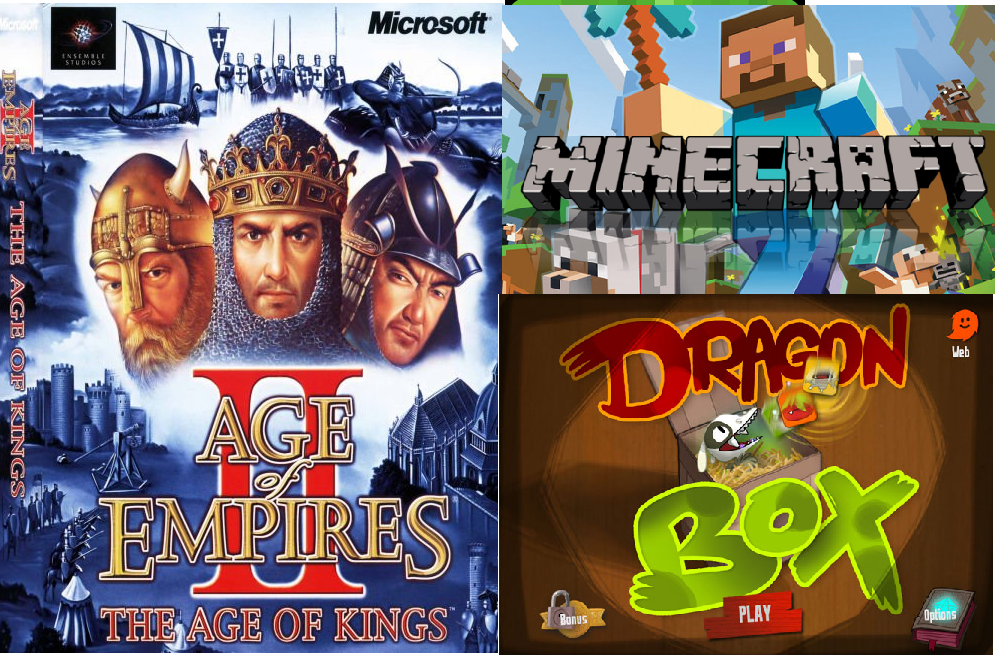}
  \caption{\textit{Exemplos de jogos usados na educação (Age of Empires II \cite{ageii}, Minecraft \cite{minecraft} e DragonBox Algebra +5 \cite{dragonbox}).}}
  \label{fig:minecraft}
\end{figure}

A maioria dos jogos educativos de sucesso hoje possuem foco em Matemática, como o \textit{Wuzzit Trouble}\footnote{\textit{Wuzzit Trouble}. BrainQuake, 2015. Disponível em: \url{http://www.brainquake.com/}.}, \textit{Slice Fractions}\footnote{\textit{Slice Fractions}. Ululab, 2014. Disponível em: \url{http://www.slicefractions.com/}.} (Fig. \ref{fig:slicefractions}) e os jogos da série \textit{DragonBox}\footnote{\textit{DragonBox}. WeWantToKnow. Disponível em: \url{http://dragonbox.com/}.} (Fig. \ref{fig:minecraft}). Eles incluem uma série de exercícios matemáticos que a criança precisa resolver para progredir no jogo. Outros jogos como \textit{Age of Empires II}\footnote{\textit{Age of Empire II: The Age of Kings}. Ensemble Studios, 1999. Disponível em: \url{https://www.ageofempires.com/}.} e \textit{Minecraft} (Fig. \ref{fig:minecraft}), que não foram desenvolvidos com o objetivo de ensinar, também são utilizados por educadores dentro de salas de aula. Analisando casos de sucesso é possível ver que a maioria deles tratam de disciplinas de história e matemática, existindo ainda grande espaço no desenvolvimento de jogos para explorar outras disciplinas.

Com isso em mente, esse projeto visa aprimorar o processo de criação e experiência de uso de jogos direcionados para crianças de 5 a 9 anos, faixa etária que compreende crianças em idade de alfabetização\footnote{A faixa etária de 6 a 8 anos de idade é tida como ideal para a alfabetização, segundo o Pacto Nacional pela Alfabetização na Idade Certa. O Pacto Nacional pela Alfabetização na Idade Certa é um compromisso formal assumido pelos governos federal, do Distrito Federal, dos estados e municípios de assegurar que todas as crianças estejam alfabetizadas até os oito anos de idade, ao final do 3º ano do ensino fundamental. \url{http://pacto.mec.gov.br/}} e que estejam próximas de iniciar este processo, como forma de auxílio no processo de ensino. Para isso, será utilizado o levantamento de requisitos do público alvo. O levantamento de requisitos é uma etapa importante para projetos de desenvolvimento de \textit{software}, e para a criação de jogos não é diferente. Ele nos ajuda a mapear as necessidades e demandas do usuário, fatores fundamentais para a criação de jogos.

Porém, ao lidar com o público infantil, essa etapa se torna mais desafiadora. É importante entender como a criança enxerga o mundo e o conteúdo didático e lúdico, para buscar o equilíbrio entre eles e as melhores formas de engajar a criança no processo de alfabetização e letramento através do jogo.

Entender o mundo da criança é fundamental para compreendermos como ela aprende e desenvolver uma ferramenta que atenda às suas necessidades e expanda seus horizontes, juntamente com as demandas dos educadores. Para tal, é preciso propor uma nova abordagem, diferenciada e própria para o universo da criança, que consiga captar as particularidades e anseios deste público.

Dessa forma, busca-se elaborar uma nova abordagem de criação de jogos educativos para o público infantil, que consiga de fato atender às demandas das crianças e facilite a implementação dos \textit{games} dentro das salas de aula com maior eficácia.

%% file: related_work.tex
A criação de jogos pode ser vista de forma similar ao desenvolvimento de \textit{softwares} e outras tecnologias. Seguindo essa lógica, os jogos devem passar por um estudo de qual é o público alvo almejado e quais os requisitos de tal público, como também a averiguação de quais funcionalidades poderiam evoluir e inovar os jogos com base nas necessidades e desejos dos usuários. Seguir os passos para o levantamento de requisitos para a criação de jogos infantis, sobretudo os educativos, é de extrema relevância para a inovação dos mesmos e de melhoria na qualidade do ensino oferecido.

\begin{figure}[h]
\centering
  \includegraphics[width=0.8\linewidth]{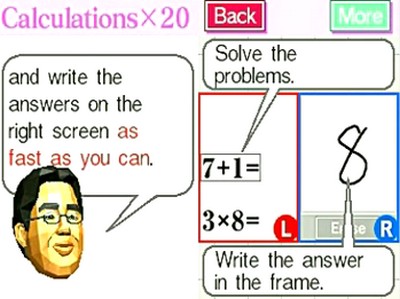}
  \caption{\textit{Jogo Dr. Kawashima's Brain Training: How Old Is Your Brain?, publicado pela Nintendo em 2005, que apresenta uma série de atividades para estimular a memória e aprendizagem. Fonte: \url{http://ds.mmgn.com/Games/Dr-Kawashimas-Brain-Training-How-Old-Is-Your-Brain}}}
  \label{fig:kawashima}
\end{figure}

Muitos jogos educativos têm sido desenvolvidos recentemente, através de incentivos governamentais, privados ou como material de pesquisa, tais como o Estrada Real Digital~\cite{oliveira2005estrada}, o jogo Super SUS, que trata sobre os direitos na saúde pública\footnote{Disponível em: \url{https://supersus.fiocruz.br/}}, e o jogo \textit{Dr. Kawashima's Brain Training} da Nintendo (Figura \ref{fig:kawashima}). Contudo, muitos desses trabalhos não têm tempo ou recursos para desenvolver melhor a etapa de levantamento de requisitos, portanto esse estudo também almeja servir de guia para outros trabalhos.

Inicialmente, as crianças eram subestimadas no processo de \textit{design} de tecnologias, no qual podemos incluir os primeiros \textit{softwares} eletrônicos com função educativa. Tais \textit{softwares} eram criados com base na ideia de exercício-e-prática ("\textit{drill-and-practice}") \cite{druin2002role}, onde as crianças tinham que responder corretamente os exercícios para avançar para o próximo módulo, e cada módulo era composto por perguntas do mesmo tema escolar. Nesse contexto, que teve início nos meados dos anos 1970 e durou até os anos 1990, podemos afirmar que as crianças eram apenas os sujeitos de pesquisa, sendo reservado exclusivamente o papel de usuário a elas, e alvos de observação.

A partir dos anos 1990, com os avanços na área de interação humano-computador e com a inovação ocorrida no mercado de jogos, a ideia da utilização de \textit{games} dentro da sala de aula foi renovada. A partir daí as crianças ganharam também um papel participativo no processo de \textit{game design}, sendo membros de times de desenvolvimento e os adultos passaram a estudar mais como o universo infantil funcionava e de que forma ocorre o aprendizado da criança. Diversos casos de participação de crianças em times de desenvolvimento são apontados por Nousiainen  (2008)~\cite{nousiainen2008exploring}, que tratam da criação de um protótipo do \textit{software} \textit{Talarius}, criado a partir de sugestões feitas pelas crianças através de desenhos de interfaces, mapas de ideias e avaliação de espaços de aprendizagem já existentes.

Ainda no começo da década de 1990, Valente (2013) ressalta a importância de \textit{softwares} educativos na educação, e divide os jogos em categorias: abordagem por instrução explícita e direta e com abordagem de exploração auto-dirigida \cite{valente1993}. Jogos educativos é um grupo da abordagem de exploração auto-dirigida e os defensores desta filosofia pedagógica acreditam que a criança aprende melhor quando é livre para descobrir relações por si só, ao invés de ser explicitamente ensinada.
Valente ainda fala em seu trabalho das categorias mais comuns de uso do computador como máquina de ensinar, sendo as mais populares os tutoriais, exercício-e-prática, jogos e simulação. É importante afirmar que as categorias trabalhadas por Valente podem se combinar e mesclar, e daí surgem os jogos de simulação ou os jogos que utilizam de exercício-e-prática para fixação de conteúdo, como o \textit{Dr. Kawashima's Brain Training: How Old Is Your Brain?}. O conceito do \textit{software} de exercício-e-prática é citado também por Druin (2002)~\cite{druin2002role}.

Já Dondlinger (2007) trata sobre elementos efetivos para o \textit{design} de jogos educativos, levantados através de uma revisão de literatura  \cite{dondlinger2007} . É apontado que até jogos de exercício-e-prática conseguem oferecer ganhos à educação, como acontece com jogos de repetição de contas matemáticas e geometria. Também são apontados outros fatores importantes para o sucesso de um jogo educativo, como a motivação e a interface do jogo. É demonstrado que jogos de aventura e estratégia são bons em motivar os alunos, sugerindo que tarefas e objetivos que exigem mais raciocínio, incluindo a procura por estratégias de resolução de problemas e tomada de decisão, agradam mais aos estudantes. Narrativas que promovem fantasia, curiosidade e desafio são inclusas como fatores que contribuem na motivação do jogador.

Magnussen et. al (2003) relatam a realização de \textit{workshops} de crianças com seu professor de matemática com o objetivo de criar um jogo sobre viagem que ensinasse matemática~\cite{magnussen2003}. Durante o processo foi pedido que as crianças sugerissem, por meio de um \textit{brainstorm}, ideias sobre o jogo. Depois disso as crianças foram divididas em grupos para trabalhar as ideias apresentadas. Contudo, esse estudo demonstrou que várias ideias apresentadas pelas crianças entravam com conflito com a didática do professor e o método de ensino.

Souza et. al  (2010) estudam a construção de jogos para apoiar o aprendizado de Pessoas com Necessidades Especiais (PNEE). É realizado um estudo etnográfico para observar a interação de grupos de alunos com jogos pré-selecionados~\cite{souza2010}. O estudo aponta requisitos importantes para esse público que possui necessidade educativa especial, como a importância da música não ser agitada, o que pode causar irritação, ou o tempo limite para a realização de uma tarefa, que pode desmotivar o aluno. Durante o trabalho foram observados fatores importantes como a facilidade do aluno em usar a \textit{interface} ou reconhecimento de personagens e objetos.

Montola  e Waern (2009)  tratam do conceito dos \textit{pervasive games}, jogos que estendem a experiência de jogo para o mundo real~\cite{montola2009}, borrando os limites do círculo mágico \cite{huizinga1971homo}, linha invisível que delimita o espaço do jogo e o diferencia da realidade. \textit{Pervasive games} são jogos que possuem uma ou mais características que expandem o círculo mágico espacialmente, temporariamente ou socialmente. Eles existem na intersecção de fenômenos tais como a cultura de uma cidade, tecnologia móvel, comunicação social, ficção e artes, combinando fragmentos de vários contextos para produzir novas experiências. O jogo educativo pode ser uma nova experiência que interliga o jogo e o ensino, que não fica contida apenas no momento em que se joga ou apenas dentro da sala de aula.

Considerando as observações e conclusões de trabalhos anteriores, neste trabalho é apresentado o processo de levantamento de requisitos preliminar para um jogo com foco em alfabetização, que analisa um grupo de alunos na faixa de 5 a 6 anos, no estado de pré-alfabetização. Na seção a seguir é relatado o desenvolvimento do trabalho e a metodologia utilizada. Deve-se ressaltar a importância do estudo de Souza et. al (2010) para trabalhos futuros, pois no Brasil o número de crianças deficientes na educação infantil tem aumentado, o que traz novos desafios para as salas de aula.

%% file: research_process.tex
Nessa seção serão apresentados alguns conceitos pertinentes ao trabalho, as decisões tomadas durante a realização da pesquisa e os métodos utilizados na mesma.

\subsection{Escolhendo os \textit{stakeholders}}

Engenharia de Software traz o conceito de \textit{stakeholder}, que são o conjunto de pessoas que têm qualquer influência ou interesse direto ou indireto sobre o projeto~\cite{project2013guide}. Existem vários stakeholders na elaboração do jogo, porém neste trabalho vamos explorar os requisitos de apenas alguns \textit{stakeholders} no processo de elaboração de um jogo infantil, através de observação e questionário: os professores, alunos e a escola. Eventualmente, durante a criação de um jogo real podem aparecer outros \textit{stakeholders}, tais como os criadores do jogo, algum investidor ou mesmo editais de governos podem exigir os próprios interesses em cima do jogo criado.

A identificação dos \textit{stakeholders} de um projeto é de extrema importância para definir todos os requisitos necessários para os diferentes grupos envolvidos. É uma forma de pensarmos na arquitetura do jogo mais completa, tentando resolver as falhas enfrentadas normalmente nos processos de desenvolvimento. Dessa forma, podemos também tratar de forma individual os interesses de cada parte envolvida, ou mesmo identificar conflitos.

\subsubsection{Aproximando a criança do processo de criação}

A primeira parte do trabalho foi identificar o público alvo da pesquisa. O público alvo foi definido como crianças de 5 a 9 anos de idade, usando como apoio as diretrizes do governo apontadas pelo Pacto Nacional pela Alfabetização Infantil, que visa alfabetizar crianças no começo do ensino básico, até os 8 anos de idade. É importante ampliar essa faixa etária no trabalho (indo de 6 anos para 5, e de 8 para 9) de forma a abranger crianças ingressando no processo de alfabetização e também aquelas que apresentarem dificuldades que levem ao atraso. Apesar do levantamento de requisitos focar em crianças na faixa etária de alfabetização, o jogo não precisa abordar apenas este conteúdo didático, também trabalhando outras disciplinas, como matemática, contação histórias e outras atividades pertinentes ao contexto desse público.

Como relatado em outros trabalhos, acontece das crianças serem deixadas de lado durante o trabalho de levantamento de requisitos, ignorando o universo infantil e focando o jogo apenas no conteúdo didático. Tal método pode acarretar em um jogo limitado que engaja pouco a criança. A participação dela no processo de \textit{design} pode ser enriquecedora, oferecendo novos ângulos na criação de jogos. Ter a oportunidade de realizar dinâmicas com as crianças e analisar suas ideias para um jogo seria o ideal para o projeto, mas devido ao tempo limitado para o trabalho, foi escolhida outra abordagem.

Procurando aproximar as crianças, foi feita uma observação naturalista de um grupo na UMEI\footnote{Unidade Municipal de Educação Infantil} Alaíde de Lisboa. A turma escolhida possuía alunos de 5 a 6 anos, em estágio de pré-alfabetização. Essa metodologia será melhor descrita na Seção \ref{sec:metodologia}. O ambiente escolar é fundamental para o trabalho porque os jogos de caráter educativo possuem o problema de conciliação do conteúdo didático com o lazer, e na sala de aula é possível ver como os alunos lidam com a aula naturalmente, procurando formas de deixar as atividades mais interessantes, sejam elas fazer um exercício de Matemática ou ler uma história.

\subsubsection{Lidando com o conflito de interesses professor-aluno}

Também foi levantado por estudos anteriores que as opiniões das crianças podem causar conflito com a natureza didática do jogo. Portanto, o que professores e alunos esperam de um jogo pode ser distinto ou até mesmo conflitante. Como o público alvo do trabalho é muito jovem e ainda há crianças na fase de pré-alfabetização, existem assuntos que eles não dominam. Além disso, um ponto sempre levantado no estudo de jogos educativos é a presença dos professores. Dessa forma, enquanto o público-alvo é a criança de 5 a 9 anos, quem faz grande parte do manuseio do jogo também é o professor, que é uma figura capaz de apontar fatores cruciais para melhorar a qualidade do material didático do jogo de uma forma que as crianças não conseguem.

Procurando contornar esses problemas, foi proposta uma outra metodologia para lidar com os professores. Foi feito um questionário para professores responderem, elucidando questões como a afinidade dos alunos com diferentes tipos de linguagens, o tempo gasto por eles na realização de tarefas, o tipo de disciplina que ensinam, por que usam jogo ou não, etc. O roteiro do questionário e seus objetivos serão explicados nas próximas seções.

\subsection{Engenharia de Requisitos}
Sommerville (2011) descreve que a área de engenharia de requisitos pode ser dividida em atividades que incluem quatro fases principais: estudo de viabilidade, elicitação (ou análise) de requisitos, especificação de requisitos e validação dos requisitos. Neste trabalho, focou-se na segunda fase, onde é feita a descoberta dos requisitos de um jogo infantil. O levantamento de requisitos engloba todas as atividades realizadas para identificar, analisar, especificar e definir as necessidades que um aplicativo deve prover para solução do problema levantado~\cite{sommerville2011engsoft}.

O levantamento e a análise compõem um processo difícil, por diversas razões \cite[cap. 4]{sommerville2011engsoft}:

\begin{enumerate}

\item Os \textit{stakeholders} frequentemente não sabem na realidade o que querem exatamente, a não ser em termos muito gerais; eles podem achar difícil articular o que desejam do jogo, por isso, podem acabar fazendo pedidos não realistas, por não terem noção do custo de suas solicitações.

\item Os \textit{stakeholders} em um sistema expressam os requisitos, às vezes, em seus próprios termos e com o conhecimento implícito de sua área de atuação. Portanto, pessoas que não possuem experiência no domínio especificado devem compreender esses requisitos.

\item Diferentes \textit{stakeholders} têm em mente diferentes requisitos e podem expressá-los de maneiras distintas. Um desafio no desenvolvimento então é durante o levantamento conseguir identificar os pontos comuns e os conflitos.

\item Alguns fatores internos podem  influenciar os requisitos do sistema. Por exemplo, alguns grupos de interesse podem definir requisitos específicos que aumentem ou diminuam a influência dele ou de outros dentro do jogo.

\item O ambiente é quase sempre dinâmico. Inevitavelmente, ele se modifica durante o processo de análise. Como consequência, a importância dos requisitos específicos pode mudar e novos requisitos podem surgir durante outras etapas do processo de desenvolvimento.
\end{enumerate}

Dados os desafios, é importante definir bem os métodos de descoberta para uma listagem mais apurada dos requisitos do jogo para que se possa transpor essas dificuldades. Durante a descoberta de requisitos várias técnicas podem ser usadas para levantá-los, como por exemplo: Levantamento orientado a pontos de vista; Entrevistas; Cenários; Casos de Uso e Etnografia (Observação Naturalista). Na próxima seção, serão listadas quais metodologias foram utilizadas neste trabalho.

\subsection{Metodologias}\label{sec:metodologia}
\subsubsection{Investigação Contextual}
O \textit{design} contextual (\textit{contextual design}) é um processo de \textit{design} da área de IHC (Interação Humano-Computador) que orienta o pesquisador a compreender profundamente as necessidades dos usuários através de uma investigação minuciosa do contexto onde ele está inserido. Essa apreciação cuidadosa do que ocorre no contexto é fundamental para o \textit{designer} elaborar uma solução adequada. As atividades do \textit{design} contextual são: investigação contextual, modelagem do trabalho, consolidação, reprojeto do trabalho, projeto do ambiente do usuário e prototipação e teste com usuários \cite[pp. 111-112]{ihc2010book}.

A observação naturalista realizada nesse trabalho consiste na etapa chamada investigação contextual (\textit{contextual inquiry}). Essa é a etapa do design contextual que lida com o levantamento de requisitos. Através dela o designer busca conhecer os usuários, suas necessidades, objetivos e como ele se comporta no seu dia a dia. A investigação ocorre no ambiente onde o usuário está inserido e onde ele realizará a utilização do produto desenvolvido. Por meio da investigação contextual se busca encontrar detalhes e motivações implícitas no trabalho dos usuários a fim de informar sobre suas necessidades reais.

A investigação contextual foi feita sobre uma turma de crianças de 5 a 6 anos na fase de pré-alfabetização na UMEI Alaíde de Lisboa. As crianças foram observadas durante o período de aula, enquanto realizavam suas tarefas escolares e outras atividades, como lanches e brincadeiras. A observação foi realizada durante aproximadamente 3 horas no período da tarde. Foi acompanhada a rotina escolar do grupo de 17 crianças, onde 10 eram meninas e 7 meninos, incluindo uma criança com necessidades especiais. Além disso a turma contava com uma professora e uma assistente, responsável pela criança com necessidades especiais.

O registro da observação foi feito através de anotações e desenhos gestuais das crianças. Foi pedida autorização prévia da coordenação da UMEI para que pudesse ser feita a observação. Não houve interação entre o observador e as crianças, e só foi conversado com a professora questões sobre a idade das crianças e em qual nível de alfabetização elas estavam.

A rotina escolar no dia consistiu em 4 atividades principais: lanche, ida à biblioteca, narrar história e atividade de aula. Na biblioteca as crianças escolheram livros para levar para casa e fazer um dever de casa. A atividade de narração consistiu em cada uma das crianças contar um trecho de uma história. Depois, as crianças engajaram em uma atividade que consistia em desenhar bolinhas que correspondessem ao número de personagens na folha de atividade, e colorir as bolinhas. Dessa forma as crianças precisavam contar os personagens e contar as bolinhas, desenhar a quantidade correta e colorir as imagens. \footnote{O relatório completo e o questionário feito para os professores podem ser acessados em \url{https://www.academia.edu/40274330/Levantamento_de_Requisitos_para_Jogos_Educativos_Infantis_-_Ap\%C3\%AAndices?source=swp_share}}

 
Ao longo da observação o foco era testemunhar comportamentos que deixassem claro quais os requisitos dos jogos seguindo as seguintes diretrizes expostas na Tabela \ref{tab:diretrizes}, \textit{guidelines} estabelecidos para o envolvimento de crianças no processo de design proposto pelo trabalho \cite{Hanna:1998:RUR:303430.303433}.

Durante o período de observação é importante frisar alguns comportamentos dos alunos:

Durante a atividade de aula:
\begin{itemize}
\item O aluno com necessidades especiais não fez a mesma tarefa que os outros, concentrando-se em brincadeiras com massinha
\item A atividade de contagem foi feita individualmente, contudo os alunos apresentaram interesse em mostrar ao colega próximo como se fazia o exercício
\item Os alunos não perguntaram em momento algum para o colega alguma coisa sobre a atividade
\item As dúvidas sobre a atividade de contagem eram perguntadas diretamente para a professora
\item Dois alunos na turma não perguntaram para a professora ou criança, mas observaram o que o colega fazia para repetir
\item Enquanto contavam e desenhavam as bolinhas apenas uma criança deixou de lado a tarefa por mais de 5 minutos
\item A maioria das crianças conseguiu se manter interessada na tarefa com poucas interrupções por aproximadamente 20 minutos
\item Enquanto faziam a tarefa as crianças faziam pequenas pausas, de menos de 2 minutos, para conversar com o colega próximo, ou conversavam sem parar de fazer a tarefa
\item A maior parte da conversa durante a tarefa era centrada em tópicos relacionados à atividade, principalmente de qual forma colorir os desenhos
\item Depois da tarefa concluída, os alunos conversavam sobre assuntos de fora da atividade
\end{itemize}
Durante a ida à biblioteca:
\begin{itemize}
\item Durante a ida à biblioteca os grupos de alunos se mostraram interessados em mostrar para os outros os livros que escolheram e as imagens que achavam legais
\item Duas crianças ficaram o tempo inteiro concentradas no livro, folheando o material, sentadas com o livro em pé, aberto e de frente para o rosto
\item Cinco crianças não apresentaram interesse nos livros
\item Os livros foram usados como objetos de brincadeiras, formando cabanas
\end{itemize}
Durante a narração da história:
\begin{itemize}
\item Todas as crianças (com exceção do aluno com necessidades especiais) demonstraram interesse em contar a história
\item Com a professora guiando, os alunos conseguiram se organizar
\item Durante a contagem de alunos, quando a criança responsável errava a contagem, uma outra criança tinha interesse em corrigir ou contar no lugar de quem errou
\end{itemize}

\begin{table}[h!]
  \small
  \begin{tabularx}{\columnwidth}{|XX|}
  \hline
  \textbf{Capacidade do aluno em realizar a tarefa dada:}                        & Qual o nível de dificuldade do jogo? Que tipo de tarefa deve ser colocada no jogo?                                  \\ \hline
  \textbf{Quando o aluno precisa de auxílio:}                                    & O aluno consegue resolver sozinho? É necessário um tutor ou um companheiro?                                         \\ \hline
  \textbf{Envolvimento do aluno na tarefa (engajamento):}                        & O aluno consegue se sentir engajado em atividades didáticas? Como?                                                  \\ \hline
  \textbf{Tempo que o aluno se mantém concentrado na tarefa:}                    & Quanto tempo o aluno consegue se concentrar numa tarefa? Que tipo de tarefa acarreta em mais ou menos concentração? \\ \hline
  \textbf{Com quais materiais o aluno tem mais afinidade:}                       & Qual será a plataforma do jogo? É possível utilizar elementos tangíveis na elaboração do jogo?                      \\ \hline
  \textbf{Como o aluno interage com os materiais:}                               & Como seria a forma de interação do aluno com interfaces diferentes?                                                 \\ \hline
  \textbf{Quais atividades atraem mais/menos o aluno:}                           & Que tipo de tarefa deve ser colocada no jogo? O que faz o jogo ser cansativo?                                       \\ \hline
  \textbf{Os alunos trabalham em grupo ou individualmente:}                      & O jogo é single player ou multiplayer? É bom jogar em grupos?                                                       \\ \hline
  \textbf{Como o aluno se comporta em grupo durante a realização de uma tarefa:} & O jogo é competitivo ou colaborativo? Como se dá a parceria entre os alunos?
  \\ \hline
  \end{tabularx}
  \caption{Diretrizes de investigação}
  \label{tab:diretrizes}
  \vspace{-0.3cm}
\end{table}
\normalsize

\subsubsection{Questionário}
Como dito anteriormente, dois \textit{stakeholders} do projeto são os professores e a escola. Com isso em mente, foi feito um questionário baseado em uma entrevista estruturada. Esta abordagem permitiu um alcance maior de professores, podendo ser respondido por um número maior de participantes, mas por outro lado, apresenta uma desvantagem por não ser possível se aprofundar tanto nos temas como em uma entrevista.

Por meio dessa entrevista estruturada se buscou elucidar: \begin{enumerate*}[label=(\roman*)] \item requisitos dos usuários primários (crianças) que não foram observados durante a investigação contextual; \item requisitos importantes para o manuseio do jogo por parte dos professores; \item requisitos sobre o material didático; \item limitações sobre a infraestrutura da escola para a utilização de jogos. \end{enumerate*} Outras questões foram descobertas após as respostas ao questionário.

%% file: others_ideas.tex
Um jeito de procurar por requisitos para o jogo educativo é encontrar o que os jogos atuais utilizados em sala de aula oferecem ou deixam de oferecer. Ver os casos já existentes e como eles são utilizados provêm uma visão sobre a demanda dos usuários e quais funcionalidades presentes nos jogos atendem as suas necessidades ou não. Para tal, foram escolhidos dois jogos que são utilizados 
com frequência em sala de aula e em casa para educar: DragonBox Algebra 5+\footnote{Recursos para aprendizado e manuais para pais e educadores disponíveis em: \url{http://dragonbox.com/educators}} e Minecraft\footnote{Versão para educação disponível em: \url{http://education.minecraft.net/}}.

\subsection{DragonBox Algebra 5+}
DragonBox Algebra 5+ é um jogo para dispositivos móveis (\textit{tablets} e \textit{smartphones}) desenvolvido para crianças a partir de 5 anos de idade, cuja proposta é ensinar álgebra. Ele é classificado como jogo educativo e possui entre 10.000 - 50.000 instalações através do Google Play\footnote{Google Play é a plataforma de downloads da Google.} e 984 resenhas, das quais 804 dão a ele 5 estrelas.

Inicialmente, como não foi possível fazer a observação das  crianças jogando, foi feita a leitura de diversos \textit{reviews} de usuários procurando por requisitos. As resenhas feitas pelos usuários revelam nas entrelinhas as dificuldades que algumas crianças têm em realizar as operações matemáticas quando o jogo avança, e como crianças que tinham dificuldade em matemática na escola conseguiram melhorar depois de jogar.
As seguintes citações exemplificam a maioria dos \textit{reviews}:

\begin{quotation}
\textit{``Truly great app. I installed this for my four and a half year old several months back. He worked his way through it almost compulsively. I think the dragons emerging from eggs is truly something that appeals to this sort of age range (or at least my son). By the end he managed to do an equation that I couldn’t do. And I have a phd in cognitive science? (although haven’t really done any equation solving for years before playing with this). The game itself is brilliantly designed. A classic in introducing complex concepts in simple visual ways. I think that properly this would be better for a slightly older child as it.''}
\\
\\
\indent \textit{``Fun for kids and grownups. I bought this game for my daughter and I ended up playing it as much as she did. She loves the game (though each puzzle takes her a lot more time) and I love seeing that she is getting the beginnings of Algebra at a young age.''}
\\
\\
\indent \textit{``Excellent for those struggling with maths. Have used this as a really good tutoring tool for algebra kids playing it outside lessons and working much better on standard equations.''}
\\
\\
\indent \textit{``Not engaging. My 6 year old got almost immediately bored its not very engaging I think other parents are right that this can be really confusing to little kids if they play it without guidance my daughter asked me bunch of questions while playing this game and finally got frustrated. Disappointing.''}
\end{quotation}
Fazendo uma análise das resenhas juntamente com o \textit{gameplay} do jogo é possível constatar pontos chaves que devem ser abordados:
\begin{itemize}
\item Elementos visuais e sonoros atraentes
\item Interface fácil de ser usada
\item Problemas desafiadores
\item Necessidade de um tutor
\end{itemize}

DragonBox Algebra 5+ é um jogo com um visual colorido e monstrinhos que, como dito por vários usuários, deixam as crianças atraídas pelos problemas matemáticos. A cada nível do jogo, a dificuldade cresce e o modo de resolução dos problemas muda, e a criança fica mais engajada pelo desafio. A tarefa deixa de ser repetitiva dessa forma.

\subsection{Minecraft}

Minecraft é um jogo que não foi designado para educação, contudo se popularizou no meio. Estudando as opiniões de professores que trabalham com ele na área da educação e as atividades propostas através da utilização do jogo, foi possível mapear as funcionalidades nele presentes que possibilitaram isso.

Existe uma comunidade \textit{online} grande de professores que utilizam Minecraft e depois de seu sucesso na área de educação foi criada uma versão Minecraft: Education Edition. Nessa versão os desenvolvedores exploram a principal característica do jogo: um mundo aberto e livre, onde o usuário tem a possibilidade de criar e ser o que quiser.

Em sua edição educacional o jogo oferece suporte aos professores para que criem mundos que podem utilizar em suas aulas, além de providenciar modelos de lições já prontos que abordam temas divididos por faixas etárias, como de 5 a 9 anos lições sobre história e de 10 a 13 anos, lições sobre preservação do meio ambiente.

Então, com Minecraft é possível enumerar pontos chaves que deixam o jogo atraente para o público jovem e para os educadores, dentre eles o principal é: a possibilidade de moldar o jogo à sala de aula, criando um ensino personalizado. Os professores possuem a possibilidade de adequar o jogo ao conteúdo didático e os alunos têm liberdade de atuação dentro do jogo.

%% file: results.tex

Após as etapas de observação, questionário e análise de outros jogos, foi possível levantar uma lista de requisitos que podemos dividir em três áreas: requisitos do aluno, requisitos do professor e requisitos da escola.

Os requisitos do aluno são o básico para que o jogo funcione. Eles tratam do tipo de interface que será criada, qual o tipo de arte será utilizada, o nivelamento da dificuldade do jogo, dentre outras coisas. Os requisitos do professor são um conjunto de requisitos que facilitam o acesso aos jogos e a utilização dentro de sala de aula. Como constatado no caso do jogo Minecraft, é importante que o professor saiba utilizar a ferramenta e ajude o aluno nisso. Os requisitos da escola falam sobre questões estruturais e de implementação: o acesso a computadores, \textit{tablets}, \textit{smartphones}, etc. 

\subsection{Requisitos do aluno}
\begin{figure}[ht]
  \centering
  \includegraphics[width=0.75\linewidth]{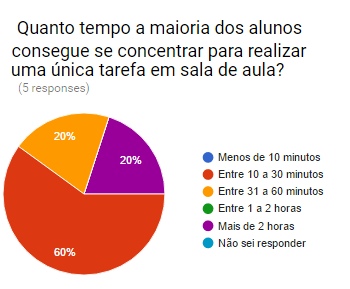}
  \caption{\textit{Tempo que as crianças conseguem se concentrar numa tarefa antes de se dispersarem}}
  \label{fig:concentracao}

  \includegraphics[width=\linewidth]{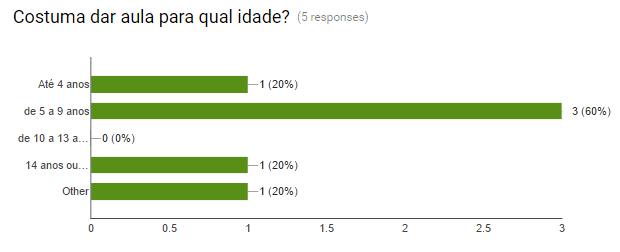}
  \vspace{-0.1cm}
  \caption{\textit{Faixas etárias dos alunos}}
  \label{fig:idade}
  \vspace{-0.2cm}
\end{figure}

Voltando às diretrizes apontadas na Seção \ref{sec:metodologia}, foram observados padrões de comportamentos importantes que se aplicam ao contexto do jogo. Foi apontado através da observação e da resposta dos professores, como mostram as Figuras \ref{fig:concentracao} e \ref{fig:idade}, que crianças de 5 a 9 conseguem se concentrar em média 20 minutos em uma única tarefa. Durante a observação foi utilizado um relógio para contar o tempo que as crianças dedicavam para as atividades e quando elas deixavam o trabalho de lado, conversando com algum colega ou fazendo outra tarefa, era considerado que haviam perdido o foco.

Também foi visto através da análise dos jogos e das atividades de colorir que as crianças se engajam em atividades didáticas, as quais normalmente rejeitariam, caso exista uma interface atraente e visual apelativo. Contudo, tarefas muito repetitivas cansam e se não há alguém para auxiliar a criança no momento de dificuldade (seja professor ou pai), ela abandona a tarefa.

A interação dos alunos com diferentes materiais demonstrou que eles gostam de desenhar, mexer com massinha, colorir, contudo, através do questionário realizado com os professores é visto que nas salas de aula a maioria das crianças na faixa etária de 5 a 9 anos se sente mais atraída por livros de história, quadrinhos e trabalhos nos computadores. O trabalho em grupo também foi observado, e segundo as perguntas feitas aos professores, as crianças preferem trabalhar em grupos pequenos.

Sendo feitas essas observações é possível listar os seguintes requisitos como base:

\begin{itemize}
\item \textbf{Atividades didáticas que não sejam repetitivas e de fácil solução}, ou então a criança pode se cansar rapidamente.

\item \textbf{Aumento da dificuldade e problemas que exigem mais tempo, mais tomada de decisões e criatividade.}

\item \textbf{Diferentes dificuldades para atender grupos diferentes de crianças.}

\item \textbf{Variedade de desafios intercalados}, de forma que quando a criança termina um tipo de atividade, ela começa outra que a estimula de forma diferente da anterior.

\item A presença de um \textbf{tutor responsável durante o uso do jogo}. As crianças gostam de desafios, mas nem sempre dão conta e é preciso de uma figura capaz de guiá-las. Seja esse o professor, seja esse um membro da família.

\item \textbf{Narrativa envolvente e contínua}: as crianças se sentem envolvidas por histórias, gostam de contar e fazer parte. Uma narrativa contínua, da qual podem fazer parte, é uma boa ferramenta de imersão.

\item O jogo com opção de texto e \textbf{narração}. Como é voltado para um público ainda em fase de alfabetização, é preciso tonar o jogo acessível às crianças que não têm plena capacidade de leitura ou que iniciaram a fase de letramento. Ouvir a narração do jogo juntamente com o texto pode ajudá-las em acompanhar o processo de alfabetização.

\item \textbf{Arte colorida e música envolvente}.

\item \textbf{Uma interface sucinta.} Jogos infantis são coloridos em sua maioria, e uma interface com muitos elementos visuais e coloridos causa confusão na criança.

\item \textbf{Suporte para jogo \textit{multiplayer} cooperativo.} Foi observado que as crianças gostam de ensinar aos colegas e mostrar novidades. Suporte cooperativo para o jogo deixaria o aprendizado mais envolvente e compartilhado. Além disso, é possível que o suporte do professor seja feito de forma \textit{on-line}, estendendo o ensino ao ambiente familiar.
\end{itemize}

\subsection{Requisitos do professor}

Os requisitos do professor tratam de questões mais importantes didaticamente, como a natureza do tema tratado pelo jogo e quais disciplinas são favorecidas pela tecnologia.

\begin{figure}
  \includegraphics[width=\linewidth]{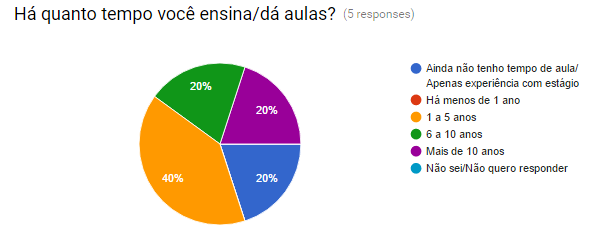}
  \caption{\textit{Tempo em que os professores participantes da segunda etapa de pesquisa dão aula}}
  \label{fig:tempoProfessor}
  
  \centering
  \includegraphics[width=0.4\linewidth]{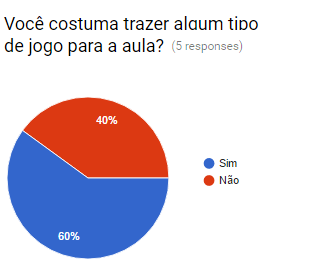}
  \includegraphics[width=0.4\linewidth]{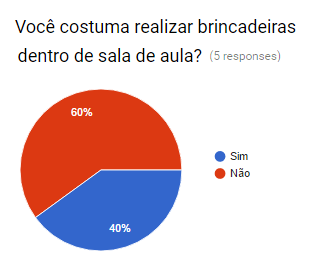}
  \caption{\textit{Proporção de professores que usam jogos e brincadeiras como ferramentas em sua sala de aula}}
  \label{fig:jogos}
\end{figure}

No questionário foram recebidas respostas de professores com diferentes tempos de trabalho, como mostra a Figura \ref{fig:tempoProfessor}. Contudo, todos comentaram como desejam poder utilizar mais jogos digitais dentro da sala de aula (Figura \ref{fig:jogos}) ou fazem jogos de outros tipos com as crianças.

Levando isso em consideração e a relevância do papel do professor como mestre da criança, foram levantados os seguintes requisitos:

\begin{itemize}

\item \textbf{Um jogo com universo expansível e maleável.} Dessa forma o professor consegue personalizar o ensino e atender às demandas da turma, além de proporcionar um material para educação contínua. Nesse cenário o professor assume o papel de mestre do mundo, podendo escolher as características do mundo e as regras.

\item \textbf{Conteúdo didático embasado nas cartilhas escolares, livros e apostilas atuais.} 

\item \textbf{Suporte aos professores}, de forma com que aprendam como usar a ferramenta para dar aula.

\item \textbf{Narrativa que enfatize questões voltadas aos valores sociais, como cultura e respeito.} 

\item \textbf{Um jogo que deva ser jogado também fora de sala de aula.} Inserir o jogo na vida da criança é uma forma de aumentar a presença e comprometimento da família na vida escolar do aluno, e atingir a realidade dele com a experiência de aprendizagem, consolidando-a. Esse é um aspecto tratado no conceito de \textit{pervasives games} que ultrapassa os limites do círculo mágico dos jogos.

\end{itemize}

\subsection{Requisitos da escola}
Nesta seção abordaremos os requisitos que surgem como demandas das escolas durante o desenvolvimento de um jogo educativo, e que foram levantados através dos questionários respondidos, questões estas que fogem do controle dos professores e alunos. 

Apesar de todos os professores que responderam ao questionário considerarem o uso de computadores importante na sala de aula, nem todas as escolas têm laboratórios de informática disponíveis para os alunos ou computadores suficientes. Essas escolas possuem uma carência em infraestrutura, o que provoca uma dificuldade em acesso aos jogos, mas também aponta para problemas de recursos financeiros.

Quando há laboratórios disponíveis, nem sempre os alunos ou professores têm acesso ao ambiente. Seja por considerarem as crianças jovens demais para o uso do computador, ou por limitações de tempo. Isso reduz as oportunidades de uso de jogos de computadores pelos alunos, existindo turmas que nunca tiveram aulas em laboratório.

Por último, as escolas fazem uso frequente de materiais prontos para auxiliar os professores em sala de aula, como apostilas, vídeos ou cartilhas. Isso demonstra a necessidade das escolas em adquirir ferramentas que sejam suplementares aos conteúdos ministrados pelos professores em sala de aula.

Observando como as crianças se relacionam com o ambiente escolar e os principais desafios enfrentados pela escola, chegamos aos seguintes requisitos:

\begin{itemize}
    \item \textbf{Criar jogos economicamente acessíveis}, que caibam no orçamento da instituição, uma vez que os recursos que a escola tem disponíveis podem ser limitados.
    \item \textbf{Desenvolver jogos que estejam disponíveis para além do ambiente escolar}, visto a restrição dos laboratórios. Jogos desenvolvidos em diferentes plataformas, como \textit{smartphones}, são uma opção para levar os jogos às crianças que não tiverem acesso a laboratórios.
    \item \textbf{Criar jogos que não exijam sessões contínuas.} Dada a subutilização de laboratórios, é necessário o desenvolvimento de um \textit{game} que possa ser aplicado em poucas sessões isoladas de aulas.
    \item Como muitos laboratórios não possuem computadores suficientes para todos os alunos, jogos com \textbf{\textit{multiplayer} local} permitem que mais de um aluno realize as atividades em um único computador.
\end{itemize}

%% file: conclusion.tex
O uso de jogos educativos tem crescido no mundo, contudo, a maioria dos jogos educativos de sucesso são encontrados e produzidos no exterior, enquanto no Brasil utilizamos versões traduzidas e adaptadas deles em sala de aula. Portanto, permanece no cenário nacional um desafio criado por questões estruturais e problemas durante o desenvolvimento de software em lidar com os requisitos de um público-alvo diferenciado: crianças em fase de alfabetização. 

Através dos requisitos levantados nesse trabalho, apontamos as peculiariedades deste público, divido em necessidades dos alunos, dos professores e das escolas do ensino público brasileiro. Observamos que os resultados envolvendo os requisitos infantis apontam para a necessidade de jogos menos repetitivos e que ocupem pequenos intervalos de tempo, de forma a assegurar a atenção do público. Também é de grande importância que as crianças tenham o auxílio de pais e responsavéis durante a execução dos jogos. Tanto para professores quanto para a escola, a infraestrutura disponível e a familiaridade dos educadores com os jogos são questões que podem ser abordadas pelos desenvolvedores ao procurarem alternativas de jogos mais acessíveis e disponibilizarem tutorais para os jogos.

Com nossos resultados, é possível preencher algumas das lacunas presentes no desenvolvimento de \textit{games} educativos e fornecer material para facilitar o caminho percorrido por novos desenvolvedores. O que foi apontado aqui pode ser usado na elaboração e efetiva aplicação de jogos de diferentes disciplinas para crianças na fase de alfabetização, procurando atender às necessidades que todos agentes envolvidos neste processo possuem.

Esse é um trabalho importante que deve ser prolongado e continuamente estudado, dando uma base maior para o desenvolvimento de jogos no Brasil e para a mudança dos métodos de ensino.

Durante a pesquisa, ficou evidente também a importância de outros fatores não abordados no escopo deste trabalho, mas que devem ser considerados em abordagens futuras, como por exemplo, os requisitos que surgem por parte dos pais e responsáveis das crianças, figuras que auxiliam nas atividades escolares, principalmente quando existem crianças com necessidades especiais, como é no caso das UMEI. Além disso, podemos aproximar ainda mais as crianças do processo de criação para aperfeiçoamento dos seus requisitos através de um acompanhamento mais prolongado e práticas de \textit{workshops} envolvendo os alunos.

Em complemento a esse estudo, também podem ser estudados quais conteúdos didáticos devem constar dentro do jogo, pois esse é um tópico que necessita de ajuda de profissionais especializados na área da educação, e não pode ser coberto apenas pelo processo de levantamento de requisitos. Pode ser feita uma pesquisa mais extensa sobre a aplicação das teorias da educação nos jogos, como o Construcionismo e o Construtivismo.

O processo de gamificação também pode ser abordado para melhorar a aplicação de jogos dentro da sala de aula, juntamente com uma análise de quais gêneros de jogos são melhores para o ensino de diferentes disciplinas. Dessa forma será possível elaborar o protótipo de um jogo educativo com foco em crianças na fase de alfabetização, o que é o intuito final desse estudo.

%% file: ms.bbl
\begin{thebibliography}{10}
\providecommand{\url}[1]{#1}
\csname url@samestyle\endcsname
\providecommand{\newblock}{\relax}
\providecommand{\bibinfo}[2]{#2}
\providecommand{\BIBentrySTDinterwordspacing}{\spaceskip=0pt\relax}
\providecommand{\BIBentryALTinterwordstretchfactor}{4}
\providecommand{\BIBentryALTinterwordspacing}{\spaceskip=\fontdimen2\font plus
\BIBentryALTinterwordstretchfactor\fontdimen3\font minus
  \fontdimen4\font\relax}
\providecommand{\BIBforeignlanguage}[2]{{%
\expandafter\ifx\csname l@#1\endcsname\relax
\typeout{** WARNING: IEEEtran.bst: No hyphenation pattern has been}%
\typeout{** loaded for the language `#1'. Using the pattern for}%
\typeout{** the default language instead.}%
\else
\language=\csname l@#1\endcsname
\fi
#2}}
\providecommand{\BIBdecl}{\relax}
\BIBdecl

\bibitem{de1998educaccao}
P.~N. de~Almeida, \emph{Educa{\c{c}}{\~a}o l{\'u}dica}.\hskip 1em plus 0.5em
  minus 0.4em\relax Edi{\c{c}}{\~o}es Loyola, 1998.

\bibitem{huizinga1971homo}
J.~Huizinga, \emph{Homo ludens: o jogo como elemento da cultura}.\hskip 1em
  plus 0.5em minus 0.4em\relax Editora da Universidade de S. Paulo, Editora
  Perspectiva, 1971, vol.~4.

\bibitem{dondlinger2007}
M.~J. Dondlinger, ``Educational video game design: A review of the
  literature,'' \emph{Journal of applied educational technology}, vol.~4,
  no.~1, pp. 21--31, 2007.

\bibitem{druin2002role}
A.~Druin, ``The role of children in the design of new technology,''
  \emph{Behaviour and information technology}, vol.~21, no.~1, pp. 1--25, 2002.

\bibitem{gros2007}
B.~Gros, ``Digital games in education: The design of games-based learning
  environments,'' \emph{Journal of Research on Technology in Education},
  vol.~40, no.~1, pp. 23--38, 2007.

\bibitem{crozat1999method}
S.~Crozat, O.~H{\^u}, and P.~Trigano, ``A method for evaluating multimedia
  learning software,'' in \emph{Multimedi a Computing and Systems, 1999. IEEE
  International Conference on}, vol.~1.\hskip 1em plus 0.5em minus 0.4em\relax
  IEEE, 1999, pp. 714--719.

\bibitem{slicefractions}
\BIBentryALTinterwordspacing
Ululab, ``Slice fractions,'' 2014. [Online]. Available:
  \url{http://pixelkin.org/2014/03/15/review-slice-fractions/}
\BIBentrySTDinterwordspacing

\bibitem{wuzzit}
\BIBentryALTinterwordspacing
BrainQuake, ``Wuzzit trouble,'' 2015. [Online]. Available:
  \url{http://www.androidpit.com.br/app/com.brainquake.wuzzittroublejr}
\BIBentrySTDinterwordspacing

\bibitem{ageii}
\BIBentryALTinterwordspacing
{Ensemble Studios}, ``Age of {E}mpires {II},'' 1999. [Online]. Available:
  \url{http://freegamecover.blogspot.com.br/2011\_07\_01\_archive.html}
\BIBentrySTDinterwordspacing

\bibitem{minecraft}
Mojang, ``Minecraft,'' 2009.

\bibitem{dragonbox}
\BIBentryALTinterwordspacing
WeWantToKnow, ``Dragon box algebra 5+,'' 2016. [Online]. Available:
  \url{https://itunes.apple.com/app/apple-store/id522069155?mt}
\BIBentrySTDinterwordspacing

\bibitem{oliveira2005estrada}
B.~D. D. S. L. T. R. M. L. C. R. F. R. C. W.~M. Alisson~Oliveira,
  Arnon~Cardoso, ``Estrada real digital,'' in \emph{Anais do WJogos-IV Workshop
  Brasileiro de Jogos e Entretenimento Digital}, 2005, pp. 242--246.

\bibitem{nousiainen2008exploring}
T.~Nousiainen and M.~Kankaanranta, ``Exploring children's requirements for
  game-based learning environments,'' \emph{Advances in Human-Computer
  Interaction}, vol. 2008, 2008.

\bibitem{valente1993}
J.~A. Valente, ``Diferentes usos do computador na educa{\c{c}}{\~a}o,''
  \emph{Em aberto}, vol.~12, no.~57, Jan-Mar 1993.

\bibitem{magnussen2003}
R.~Magnussen, M.~Misfeldt, and T.~Buch, ``Participatory design and opposing
  interests in development of educational computer games,'' \emph{Level Up},
  2003.

\bibitem{souza2010}
F.~de~Souza, M.~M.~V. Paula, A.~C.~B. Ramos, and M.~M. Souza, ``Necessidades
  educativas especiais: constru{\c{c}}{\~a}o de jogos para apoiar o
  aprendizado,'' \emph{Proceedings do IX Brazilian Symposium on Games and
  Digital Entertainment - SBGames}, vol.~9, 2010.

\bibitem{montola2009}
M.~Montola, J.~Stenros, and A.~Waern, \emph{Pervasive Games: Theory and
  Design}.\hskip 1em plus 0.5em minus 0.4em\relax San Francisco, CA, USA:
  Morgan Kaufmann Publishers Inc., 2009.

\bibitem{project2013guide}
P.~M. Institute, \emph{A Guide to the Project Management Body of Knowledge:
  PMBOK Guide}, ser. PMI Standard.\hskip 1em plus 0.5em minus 0.4em\relax
  Project Management Institute, 2013.

\bibitem{sommerville2011engsoft}
I.~Sommerville, \emph{Engenharia de software}.\hskip 1em plus 0.5em minus
  0.4em\relax Pearson Education do Brasil, 2011, vol.~9.

\bibitem{ihc2010book}
S.~D.~J. Barbosa and B.~S. da~Silva, \emph{Intera{\c{c}}{\~a}o
  Humano-Computador}.\hskip 1em plus 0.5em minus 0.4em\relax Elsevier Brasil,
  2010.

\bibitem{Hanna:1998:RUR:303430.303433}
L.~Hanna, K.~Risden, M.~Czerwinski, and K.~J. Alexander, ``The design of
  children's technology,'' A.~Druin, Ed.\hskip 1em plus 0.5em minus 0.4em\relax
  San Francisco, CA, USA: Morgan Kaufmann Publishers Inc., 1998, ch. The Role
  of Usability Research in Designing Children's Computer Products, pp. 3--26.

\end{thebibliography}
